The effect of primary school education on preventive behaviours during COVID-19 in Japan

Short title: Education and preventive behaviours

Eiji Yamamura[1*], Yoshiro Tsutsui[2], Fumio Ohtake[3],

[1] Department of Economics, Seinan Gakuin University, Fukuoka, Japan

*Corresponding author
Email: yamaei@seinan-gu.ac.jp (EY)
A full list of author information is available at the end of the article

# Abstract

## Background

Education plays a critical role on promoting preventive behaviours against the spread of pandemics. In Japan, hand-washing education in primary schools was positively correlated with preventive behaviours against COVID-19 transmission for adults in 2020 during the early stages of COVID-19 [1]. The following year, the Tokyo Olympics were held in Japan, and a state of emergency was declared several times. Public perceptions of and risks associated with the pandemic changed drastically with the emergence of COVID-19 vaccines. We re-examine whether effect of hand-washing education on preventive behaviours persisted by covering a longer period of the COVID-19 pandemic than previous studies.

## Methods

26 surveys were conducted nearly once a month for 30 months from March 2020 (the early stage of COVID-19) to September 2022 in Japan. By corresponding with the same individuals across surveys, we comprehensively gathered data on preventive behaviours during this period. In addition, we asked about hand-washing education they had received in their primary school. We used the data to investigate how and the degree to which school education is associated with pandemic mitigating preventive behaviours.

## Results

We found that hand-washing education in primary school is positively associated with behaviours such as hand washing and mask wearing as a COVID-19 preventive measure, but not related to staying at home. We observed a statistically significant difference in hand washing between adults who received childhood hand-washing education and those who did not. This difference persisted throughout the study period. In comparison, the difference in mask wearing between the two groups was smaller, but still statistically significant. Furthermore, there was no difference in staying at home between them.

## Conclusion

Childhood hygiene education has resulted in individuals engaging in hand washing and mask wearing to cope with COVID-19. Individuals can form sustainable development-related habits through childhood education.

# Introduction

Ideally, individuals and organisations would have better preparedness for unexpected events such as pandemics prior to their emergence. Education, which can improve preparedness, is expected to alter individuals' risk perceptions and positively affect health outcomes. Educational level played critical role in an individual's preparedness for the COVID-19 pandemic [2]. Risk perception and precautionary behaviour against pandemics can be dynamic over time [3]. However, the effect of education persists for a longer period once hygiene habits are formed [1,4], contributing to the sustainability of society.

Individuals can cope more smoothly to the occurrence of epidemics and pandemics if they take appropriate precautions. According to Ikeda et al. [5], individuals in Japan care about hygiene, such as regular hand washing, reducing the risk of contracting an infection. Lee et al. explored the role of hygiene education of children in schools on regular hand-washing behaviours during the COVID-19 pandemic [1]. They found that hand-washing education in primary school is positively correlated with various preventive behaviours in adulthood during the COVID-19 pandemic but also prior to the pandemic. This finding was observed using a dataset collected from April to August 2020 during the early stage of the COVID-19 pandemic.

According to a study on the 2009 H1N1 influenza pandemic, the Mexican government promoted campaigns to educate the public about using hand sanitisers, hand-washing techniques, and wearing masks. Accordingly, past experiences with pandemics facilitated hand-washing behaviours, which provided long-lasting effects on health outcomes [4]. The COVID-19 pandemic continued to influence various aspects of daily life until at least the end of 2022. Individuals learned about COVID-19 based on their experiences and public campaigns. However, COVID-19 vaccination was implemented in 2021 throughout Japan, and most individuals have been vaccinated. Hence, the risk of COVID-19 infection is reduced, which influences the degree of engagement in preventive behaviours [6–8].

Overall, the situation changed drastically from the early stages of the COVID-19 pandemic. It is necessary to consider how and the extent to which the effect of childhood education on preventive behaviours changed. In this short note, we used monthly individual-level longitudinal data to re-examine the findings of Lee et al. over a longer time period [1]. We asked: How did hygiene practice education in childhood influence preventive behaviours in adulthood during the COVID-19 pandemic from March 2020 to September 2022?

## Materials and methods

### Data collection

Shortly after COVID-19 infection was detected in Japan in January 2020, we decided to collect data through online surveys by commissioning the research company INTAGE. INTAGE was chosen for their good reputation due to their abundant experience with academic research. In the early stages of the COVID-19 pandemic (March 13–16, 2020), the first wave of queries was conducted to gather 4,359 observations. INTAGE recruited participants for the survey from among pre-registered individuals, with a participation rate of 54.7 %. Respondents were randomly selected to fill the pre-specified quotas by identifying a representative of the Japanese adult population (ages 18–78 years), and data was collected on household income, age, gender, educational background, and area of residence. This sampling method was chosen because individuals aged 17 years and below were too young to be registered with INTAGE, and data from individuals over 78 years of age could not be reliably collected mainly because they were unlikely to use the Internet. Consequently, the sample population was restricted to ages 18–78 years.

Longitudinal panel data were constructed as follows. Internet surveys were conducted nearly monthly on 26 occasions ('waves') between March 2020 and September 2022 with the same individuals. Surveys were not conducted for three months between July-September 2020 because of a shortage of research funds. After acquiring additional funding, surveys continued in October 2020 (6$^{th}$ wave) and included an additional question on primary school education to examine the effect of childhood education on preventive measures. The first survey by Lee et al. was conducted between April 28–30 [1]. By comparison, we conducted our first survey one month earlier, between March 13–27. From March to April 2020, the COVID-19 situation in Japan changed drastically, making this a notable distinction [9–12]. During the study period, some respondents stopped taking the surveys and were removed from the sample pool. We limited samples used for analysis to respondents who participated from the first to the 26$^{th}$ wave to follow the same individuals. Further, we restricted the sample to those who answered various questions, such as primary household income, job type, and education in primary school. In particular, many respondents did not remember experiencing hand-washing education in primary school. Eventually, the number of respondents was reduced to 996, and the total number of observations used in this study was 25,896.

## Methods

Table 1 describes the key variables used in the estimation and reports their means and standard deviations. The survey questionnaire contained basic questions about demographics, such as birth year, gender, educational background, household income, and jobs.

**Table 1. Definitions of key variables.**

| Variables | Definition | Mean | s.d. |
|---|---|---|---|
| | Outcome variables | | |
| STAYING HOME | In the last week, how consistent were you at 'not going out of home'? Please choose among 5 choices. 1 (not completed at all) to 5 (completely consistent). | 4.21 | 0.91 |
| WEARING MASK | In the last week, how consistent were you at 'wearing a mask'? Please choose among 5 choices. 1 (not completed at all) to 5 (completely achieved). | 4.54 | 0.19 |
| HAND WASHING | In the last week, how consistent were you at 'washing your hands'? Please choose among 5 choices. 1 (not completed at all) to 5 (completely achieved). | 2.91 | 1.29 |
| | Confounders (Independent variables) | | |
| WASHING EDUCATION | Did everyone in your class was supervised by teachers to ensure that they washed their hands in turn? 1 (Yes) or 0 (No) | 0.48 | 0.49 |
| SCHOOL UNIFORM | Did you wear school uniforms in primary school? 1 (Yes) or 0 (No) | 0.20 | 0.40 |

The estimated function takes the following form:

$$Y_{it} = \alpha_0 + \alpha_1 \text{ WASHING EDUCATION}_{it} + \alpha_2 \text{ SCHOOL UNIFORM}_{it} + k_t + u_{it}$$

$Y_{it}$ is the outcome variable for individual i and wave t and α denotes the regression parameters. $u_{it}$ is the error term. The estimation method was the ordinary least squares model. The behaviour of individuals depends on the situation. For instance, residents were strongly requested to stay at home during states of emergency. There were also cycles of increasing and decreasing numbers of new infections, which were common in all parts of Japan [11,12]. $k_t$ represents the characteristics of the situation at each time point. To control for this, we used 25 time point dummies.

Y is the outcome variable captured by the three proxy variables STAYING HOME, HAND WASHING, and WEARING MASK. The respondents were asked the following questions about preventive behaviours:

'Within a week, to what degree have you practiced the following behaviours? Please answer based on a scale of 1 (I have not practiced this behaviour at all) to 5 (I have completely practiced this behaviour)'.

(1) Staying home

(2) Wearing a mask

(3) Washing my hands thoroughly

The answers to these questions served as proxies for the following variables for preventive behaviours: staying home, frequency of hand washing, and degree of mask wearing. Larger values indicate that respondents are more likely to engage in preventive behaviours.

The key confounding variable is WASHING EDUCATION, which is '1' if teachers supervised pupils to ensure that they washed their hands during primary school, otherwise it is '0'. In this study 48% of respondents had experienced hand-washing practice in primary school (Table 1). Previous studies have found that the experience of wearing school uniforms during primary school is positively correlated with pro-social inclinations in adulthood [13]. Therefore, the experience of school uniforms may be correlated with preventive behaviours in adulthood, so SCHOOL UNIFORM is also included as a confounding variable. The statistical software used in this study was Stata/MP 15.0.

# Results

## Baseline estimations

**Table 2. Dependent variables are preventive behaviours (Data: 1st to 26th wave).**

|  | (1) HAND WASHING | (2) WEARING MASK | (3) STAYING HOME |
|---|---|---|---|
| WASHING EDUCATION | 0.197*** (0.126-0.267) | 0.109*** (0.051-0.167) | 0.072 (−0.032-0.186) |
| SCHOOL UNIFORM | 0.026 (−0.067- 0.119) | 0.026 (−0.065-0.061) | −0.002 (−0.193-0.100) |
| *Time Fixed Effects* | Yes | Yes | Yes |
| *Control variables* | Yes | Yes | Yes |
| Adj $R^2$ | 0.11 | 0.19 | 0.14 |
| Obs. | 25,896 | 25,896 | 25,896 |

**Note**: Numbers without parentheses are coefficients of the confounding variables. Numbers within parentheses are 95% CI. The model includes various control variables, such as age, gender, dummies for household income, job dummies, number of deaths, and number of infected people in residential prefectures. However, these results have not been reported. "Yes" means that these variables are included.

***p<0.01

Table 2 shows the estimation results and coefficient of confounders. We found a positive correlation for WASHING EDUCATION for all dependent variables. The relationship between WASHING EDUCATION and both HAND WASHING and WEARING MASK were found to be statistically significant, whereas STAYING HOME is not. The coefficient of HAND WASHING is 0.198, meaning that those who experienced hand-washing education in primary school are more likely to wash their hands by 1.987 points on a 5-point scale compared to those who did not. The effect of hand-washing education on HAND WASHING was approximately two times larger than that of WEARING MASK (0.109). We did not find statistical significance for SCHOOL UNIFORM on any dependent variable.

Overall, hand-washing education in childhood promotes the hygiene practice of hand washing and wearing masks, but did not promote staying at home. Table 3 shows that the results of waves 1–5 are almost identical to those in Table 2.

**Table 3. Dependent variables are preventive behaviours (Data: 1st to 5th wave).**

|  | (1) HAND WASHING | (2) WEARING MASK | (3) STAYING HOME |
|---|---|---|---|
| WASHING EDUCATION | 0.196*** | 0.121* | 0.037 |
|  | (0.103-0.289) | (−0.008-0.251) | (−0.069-0.136) |
| SCHOOL UNIFORM | −0.029 | −0.069 | −0.013 |
|  | (−0.127- 0.069) | (−0.190-0.051) | (−0.135-0.109) |
| *Time Fixed Effects* | Yes | Yes | Yes |
| *Control variables* | Yes | Yes | Yes |
| Adj $R^2$ | 0.11 | 0.17 | 0.14 |
| Obs. | 4,980 | 4,980 | 4,980 |

**Note**: Numbers without parentheses are coefficients of the confounding variables. Numbers within parentheses are 95% CI. The model includes various control variables, such as age, gender, dummies for household income, job dummies, number of deaths, and infected persons in residential prefectures. However, these results have not been reported. "Yes" means that these variables are included.

*p<0.10

***p<0.01

## Changes of preventive behaviours

Figs 1–3 show that preventive behaviours drastically increased during the early stage

of COVID-19, especially during the first state of emergency, as indicated by the solid vertical line. Subsequently, in Figs 1 and 2, hand washing and mask wearing were maintained at high levels throughout the study period. This is in contrast with the findings in Japan that precautionary behaviour in response to the 2009 (H1N1) influenza pandemic fluctuated [3].

Those who experienced hand-washing practices in primary school were more likely to engage in hand washing and mask wearing during the COVID-19 pandemic. The effect of hand-washing education on mask wearing was smaller and less statistically significant than that on hand-washing. This may be because hand-washing education is more likely to form a habit of washing hands than wearing masks. Wearing masks in crowded places is effective in mitigating pandemics, whereas wearing masks in open air is much less effective [14]. People wear masks outdoors, partly because of peer pressure.

Hand-washing education played a critical role in forming lasting habits of health-protective behaviours such as hands-washing and mask wearing. By contrast, Figure 3 shows the fluctuating cycles of staying at home. Furthermore people became overall less likely to stay home after the COVID-19 vaccine was implemented, as indicated by the dashed vertical line. People are unlikely to form a habit of staying home, which is congruent with Ibuka et al. [3]. There was no difference in staying home between those who had experienced hygiene education practice and those who did not.

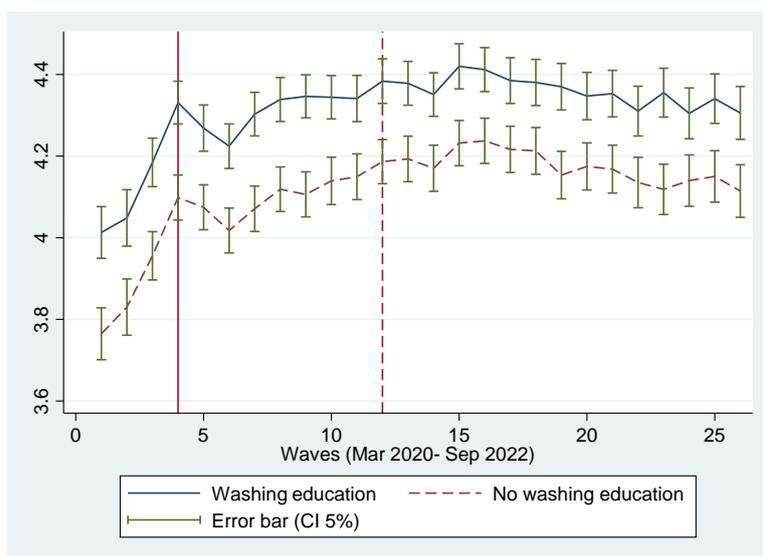

Fig 1. Hand-washing behaviour

Note: The solid vertical line indicates when the first state of emergency was declared in Japan. The dashed vertical line shows when COVID-19 vaccination was implemented.

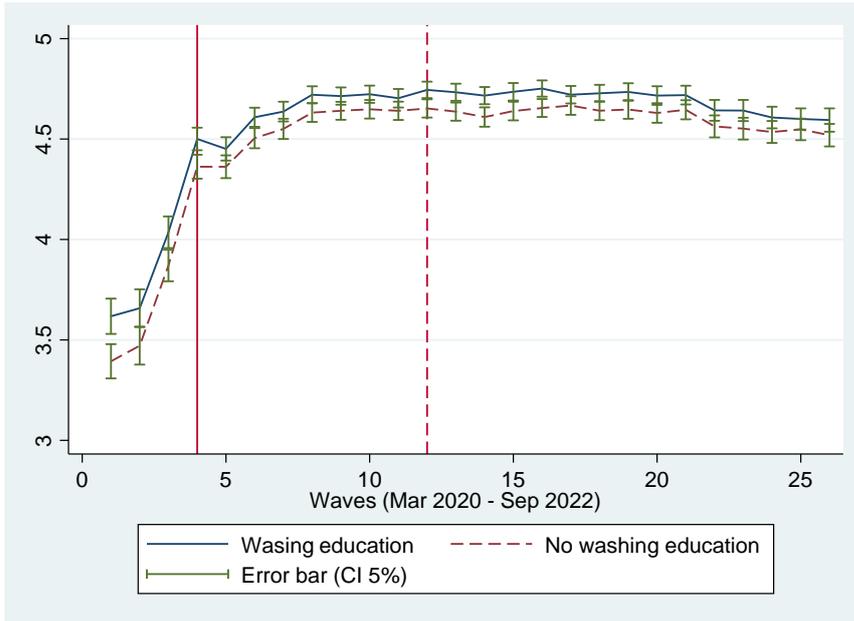

Fig 2. Mask-wearing behaviour

Note: The solid vertical line indicates when the first state of emergency was declared in Japan. The dashed vertical line shows when COVID-19 vaccination was implemented.

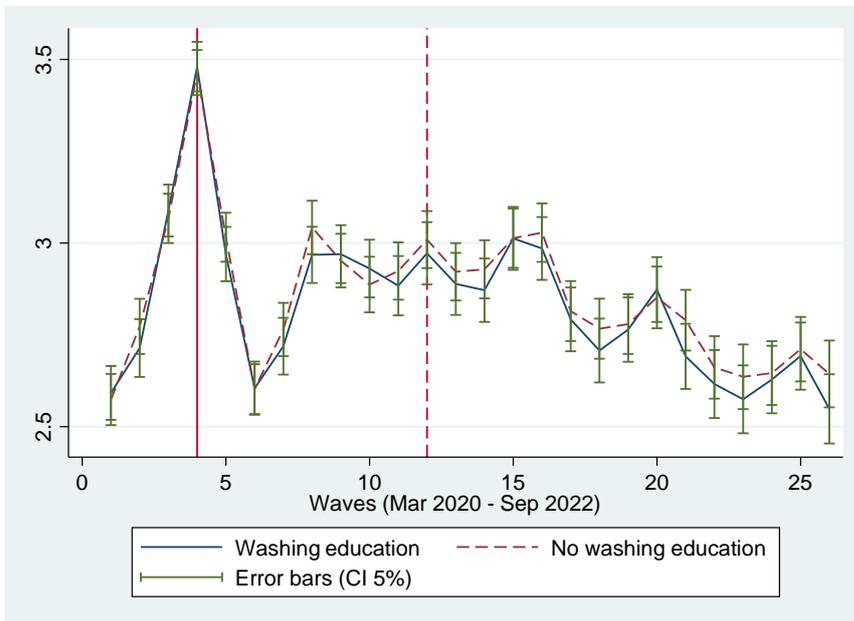

Fig 3. Staying at home behaviour

Note: The solid vertical line indicates when the first state of emergency was declared in Japan. The dashed vertical line shows when COVID-19 vaccination was implemented.

# Discussion

The purpose of this study is to consider how school practices in primary schools influenced preventive behaviours during the COVID-19 pandemic using data covering March 2020 to September 2022. Preventive behaviours reduce one's own risk of being infected, but also the risk of infecting others. Therefore, preventive behaviours against pandemic spread can also be considered an investment in public goods to benefit society [15]. Lee et al. found that hand washing led people to display preventive behaviours even before COVID-19 (Appendix 7) [1]. Considering their and our findings together, hygiene education resulted in a habit of hygiene preventive behaviours and persisted regardless of pandemic severity.

Lee et al. [1] found that hand-washing education is positively associated with various preventive behaviours, including wearing masks and staying at home. In contrast to Lee et al., this study found clear differences in educational impact according to the type of preventive behaviour. In the questionnaire used by Lee et al., detailed questions about primary school education and various preventive behaviours were included. The respondents may have perceived the researchers' intentions, which may have influenced their responses. For example, their questions about hygiene practice in primary school may have functioned as a 'nudge' that unintentionally influenced respondents to meet the goals of the researchers and respond accordingly. In our study, questions about preventive behaviours were included in all waves, whereas questions about primary school education were blended into various questions only in the 6th wave onward. Hence, before the 6th wave, respondents may not have perceived our goals to associate childhood education with preventive behaviours. In order to directly compare our results with those of Lee et al., we analysed data from the 1st through 5th waves which are almost equivalent to the period they studied, conducting estimations using the same specifications (Table 3).

Staying at home was not significantly correlated with hand-washing education during childhood. This might be because staying at home is a different type of preventive behaviour than hand washing. People stay home only when their benefits exceed their costs. People sacrifice various experiences through outdoor activities in the real world if they stay at home. In economic terms, this sacrifice is considered an 'opportunity cost' of staying home. As the opportunity cost is not reduced even if one experiences hand-washing education in childhood, individuals will stay at home only if their benefits outweigh their costs regardless of hygiene education. Additionally, staying at home weakens social ties and reduces social capital because of a reduction in social interaction through face-to-face communication. As is widely acknowledged, social ties and social capital are positively associated with health status [16–18]. Therefore, it is important to distinguish staying at home from other preventive behaviours.

The formation of hand-washing habits through hygiene education in childhood reduces its psychological costs. In this case, people do not need to change their lifestyle to engage in basic preventive behaviours such as hand washing, regardless of the severity of the pandemic. Basic hygiene practices in childhood have reduced stress in life during the pandemic.

## Strength

We constructed longitudinal data to cover a longer period than previous studies in Japan, where preventive behaviours were not enforced with penalties[1,3]. Lee et al. did not examine the effects of the emergence and spread of the COVID-19 vaccine on preventive behaviours [1]. Preventive behaviours of individuals were thought to change in response to the emergence of the COVID-19 vaccine. However, individuals continue to wash their hands and wear masks long after vaccine implementation. This clearly suggests these preventive behaviours are stable.

## Limitation

Many respondents did not remember experiencing hand-washing education in primary school. We have deleted them from the data pool used for analysis. There was a difference in the characteristics of respondents who answered the questionnaire and those who did not. This may have resulted in selection biases. Furthermore, answers to the questionnaire seem to depend not only on the facts, but also on the respondent's misapprehension. Therefore, recall bias may occur. Another variable of school education would show statistical significance if biases had a significant effect on the results. However, SCHOOL UNIFORM is not significantly correlated with preventive behaviours, which is clearly different from the results of WASHING EDUCATION. This suggests, to a certain extent, the biases are minor.

Wearing masks are less effective in open air than indoors [14]. In contrast to Lee et al. [1], we used only three proxies for preventive behaviours. Therefore, we did not scrutinise how hand washing and mask wearing changed in different situations.

In contrast to hand washing, the benefit of mask wearing depends on the situation. Wearing masks in the open air has limited effectiveness [14]. In mid-summer, wearing masks increased the risk of heatstroke. In this situation, the cost of wearing a mask is higher than its benefits. It is, therefore, important to examine mask-wearing behaviour in various situations in future studies.

## Conclusion

Preventive behaviours play a vital role in coping with unexpected pandemics such as COVID-19. We concluded that people can form sustainable development-related habits through childhood hygiene practice education.

## Acknowledgements

We would like to thank Editage (http://www.editage.com) for their English Language editing and reviewing of this manuscript.

## Authors' contributions

EY and FO participated in the conceptualisation of the study and analysed the patient data. YT designed the panel survey and performed data collection. EY wrote the main text and made the tables for the original manuscript. All authors reviewed, edited, and approved the final manuscript. The authors are responsible for any errors in this study.

## Funding

This study was supported by the Fostering Joint International Research B (Grant No. 18KK0048) and the Grant-in-Aid for Scientific Research S (Grant No. 20H05632) from the Japan Society for the Promotion of Science to Yoshiro Tsutsui and Fumio Ohtake, respectively.

## Availability of data and materials

The datasets used and analysed in this study are available from the corresponding author upon reasonable request.

## Declarations

## Ethics approval and consent to participate

This study was conducted with the ex-ante approval of the Ethics Committee of the Graduate School of Economics, Osaka University, and all methods were carried out in accordance with the relevant guidelines and regulations. The ethics approval number of Osaka University for this study is R021014. Informed consent for study participation was obtained from all subjects.

## Consent for publication

Not applicable.

## Competing interests

The authors declare that they have no competing interests.

Author details
1. Department of Economics, Seinan Gakuin University, Fukuoka, Japan. 2. Faculty of Social Relations, Kyoto Bunkyo University, Kyoto, JAPAN, 3. Center for Infectious Disease Education and Research (CiDER), Osaka University, Osaka, Japan.